\documentclass[twocolumn,showpacs,preprintnumbers,amsmath,amssymb]{revtex4}
\usepackage{epsfig}

\newcommand{\Vr}{{\bf r}}

\begin{document}
\draft

\title{Scaling of polymers in aligned rods}

\author{J. M. Deutsch and M. Warkentin}
\affiliation{
Department of Physics, University of California, Santa Cruz, CA 95064.}

\date{\today}

\begin{abstract}
We study the behavior of self avoiding polymers in a background of
vertically aligned rods that are either frozen into random positions or free
to move horizontally. We find that in both cases the polymer chains
are highly elongated, with vertical and horizontal size exponents that differ
by a factor of 3. Though these results are different than previous
predictions, our results are confirmed by detailed computer simulations.
\end{abstract}


\maketitle 

The statistics of a flexible polymer molecule in a background of randomly
placed vertical rods has been considered previously~\cite{baumgartner,dimarzio}.
The problem was first investigated~\cite{baumgartner} in a dynamical context in order
to better understand the effects of entanglements. Later,~\cite{dimarzio} the
equilibrium statistics were examined, and an elongation of the chain in the
vertical direction was predicted. As we will show using both scaling and 
simulations, elongation does take place, but is actually much larger than was 
first thought.

We also analyze how this problem is very closely related to a polymer in
a background of {\em mobile} vertical rods. We argue that this would
be an interesting system to investigate experimentally, perhaps being
related to liquid crystal polymer mixtures.

This work has overlap with many other problems of interest. For example,
vortices in superconductors with columnar disorder has been extensively
studied~\cite{nelson,nattermann} and is related to the current polymer problem, a major
difference being that the polymer (or vortex) was stretched and threaded the 
entire vertical dimension having a line tension, meaning that it is
appropriate to use directed polymers in this situation~\cite{kardar}. This apparently leads to
completely different scaling behavior~\cite{arsenin}.

Our findings are also relevant to the study of mixtures of rods and
polymers. It is possible to solubilize many rod-shaped particles---for example
nanotubes~\cite{nanotubes}, or more commonly, virus particles,  boehmite rods,
and cellulose nanocrystals---and
rod-polymer mixtures have been studied experimentally and theoretically
in the context of demixing and liquid crystalline phase
transitions~\cite{dogic,sear,buitenhuis,vanbruggen,edgar,tang,devries}.
Furthermore, it is now possible to possible to fabricate arrays of 
vertical nanotubes on a substrate~\cite{semet} which allows for the possibility of 
observing the scaling behavior predicted here (the case of quenched 
disorder).

We use a three dimensional cubic lattice model to study this problem,
where the rods randomly occupy vertical lines, and the chain avoids itself
and the rods. This is related to the problem of self avoiding chain (SAW)
that also is excluded from {\em point} defects. In that case, an elegant
argument of Cates and Ball~\cite{catesball} cleared up a decade of controversy
by showing that the quenched version
of this problem, where defects are frozen, gives exactly the same statistics
for the SAW as the annealed version, where the defects are also mobile.
They argued that a frozen background of uncorrelated obstacles on an infinite lattice
could be subdivided into very large regions. The statistics of obstacles in
each region are independent of each other. Therefore the statistics of the
polymer chain can be obtained by doing an average over all these regions. The
statistical weight given to each region gives precisely the same result as
an annealed average.  Their argument trivially extends to the case we consider
here.

From the above paragraph, we conclude that for an infinite sized lattice, the
problem of frozen rods gives identical polymer statistics to the case where the rods
are mobile. Below we will analyze how finite size lattices alter the
above conclusion. 
But first we analyze the annealed (mobile rod) problem.

Denote the probability distribution of the SAW with coordinates $\Vr_1,\Vr_2,\dots,\Vr_N$,
as $P_{SAW}\{\Vr_i\}$. When it is placed in a random potential $V(\Vr)$ at
temperature $T$, the annealed average probability distribution becomes 
\begin{equation}
\label{eq:anneal}
P_{SAW}\{\Vr_i\} \left< e^{-{1\over T}\sum_j V(\Vr_j)}\right>_V
\end{equation}

The average depends on the statistics of $V$. First we review the case
of a completely uncorrelated $V$~\cite{catesball,machta}, in which
case the average becomes an on-site attractive term between different
monomers.  However because the SAW excludes all configurations where
two monomers sit on the same site, this cannot alter the probability
distribution. Therefore an uncorrelated random potential makes {\em no
difference} to the statistics of an SAW. This result has been confirmed
by simulations~\cite{lee}.

Now we turn to the case of columnar disorder. Here the average in eqn \ref{eq:anneal} 
now leads to an attractive interaction of all monomers that
have the same x and y coordinates, regardless of their separation in the z direction.
In this case, we expect different scaling behavior than we would observe from an SAW. Intuitively,
the chain should contract in the x-y plane. We now analyze the annealed statistical
mechanics of this problem using a scaling argument.

A dilute gas of rods of density (per unit area) of $\rho$ and at temperature $T$ gives rise to a 
(two dimensional) pressure $p = \rho T$. An SAW  of $N$ steps placed in a gas of these rods will raise
the free energy of the rods by excluding rods from the vicinity of the SAW.
This is due to the pressure exerted by the rods which means that the free
energy is raised by $p A$ where $A$ is the cross-sectional area of the rods
in the $x-y$ plane. Therefore the SAW has forces acting on it to decrease $A$.
Working in opposition to this is the entropy loss of a polymer chain confined to a
cylinder of radius $R_{xy}$ which is~\cite{degennes}  
$\sim N/N_{xy}$ where $N_{xy}$ is the number of monomers corresponding to an SAW
of dimension $R_{xy} \sim N_{xy}^\nu$, where $\nu \approx .59$ is the excluded volume 
exponent in three dimensions. Minimizing the pressure term and tube confinement terms gives:
\begin{equation}
\label{eq:rxy}
R_{xy} \sim N^{1/(2+1/\nu)} \approx N^{0.27} .
\end{equation}

Note that the exponent $0.27$ is much less than in the pure SAW case---the pure SAW problem is isotropic and so $\nu$ is as well. This shows that the scaling of the polymer
chain in this situation differs greatly from a pure SAW. To obtain scaling in the vertical
direction $R_z$, we note that there are ``blobs" of size $R_{xy}$ stacked vertically on top
of each other. Therefore 

\begin{equation}
\label{eq:rz}
R_z \sim R_{xy} N/N_{xy} \sim  N^{3/(2+1/\nu)}
\sim (R_{xy})^3 \approx N^{.81} .
\end{equation}

This exponent is thrice the $x-y$ exponent, and is larger than $\nu$ for the pure SAW.
Therefore for long chains the system becomes highly elongated. The
overall radius of gyration therefore scales the same way as $R_z$.

Note that for long chains, the chain density projected on to the x-y
plane grows algebraically for large $N$. Therefore for large
$N$ we expect complete exclusion of rods in this region, which is
consistent with our initial assumption that this was the case. In addition
because monomers in far away blobs are uncorrelated, and the number of
blobs grows with $N$, single configurations
all have very close projected x-y densities which will only differ
from each other by an inverse power of N. This self-averaging
property implies that we are also justified in regarding the
polymer as being in an external potential, induced by the rods,
and then minimizing the free energy to find the scale of this
potential. Therefore we expect our scaling argument to work very
well for large $N$.

As argued above, this is {\em identical} to the scaling expected in 
the case of a frozen background for a system of infinite size. However
the case of frozen finite size systems is harder to analyze through analytical means
as the above argument does not apply. In general for a finite
size system, one does expect a difference between the quenched and annealed
cases.
Therefore we performed extensive simulations of this system to verify
the above predictions and better understand the effects of finite
sizes.

Chains of length 32 to 4096 were
simulated on a cubic lattice with different densities of vertical rods. First we allowed both rods
and chains to move according to rules that satisfied detailed balance.
The polymers were moved according to reptation dynamics~\cite{wallmandell} and the rods performed
long range moves 
between randomly selected positions also satisfying detailed balance.
We checked that our program did indeed give a reasonable excluded
volume exponent in the case of no rods, $\nu \approx .59$ and performed
a variety of other checks.

\begin{figure}
\centerline{\epsfxsize=\columnwidth \epsfbox{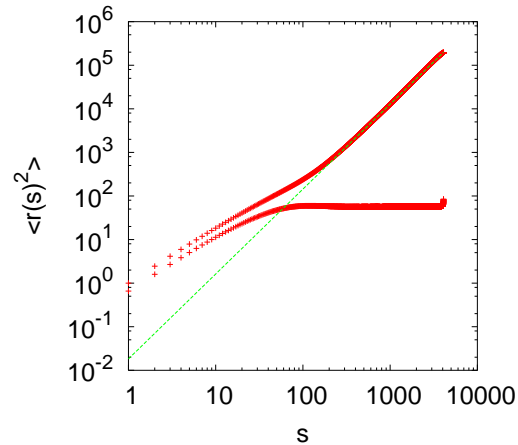}}
\caption
{
Mean monomer separation squared as a function of arclength defined in eqn \ref{eq:rsq}, both the total (upper curve)  
and the 
x-y projection (lower curve) for an SAW of 4096 steps and vertical rods with
$\rho = 0.32$ with mobile rods. The best fit to the upper graph is
line is $2\times .974$.
}
\label{fig:rsq}
\end{figure}

We then analyzed the statistics of
our longest chain and highest rod density, $N=4096$ and $\rho = .32$.  We 
plot the statistics for the mean monomer separation squared as a function of arclength
and the same thing but just projected in the x-y plane, that is
\begin{eqnarray}
\label{eq:rsq}
\left<r^2(s)\right> \equiv \sum_{i<j} \delta_{j-i,s} {1\over N-s} \left< |\Vr_i-\Vr_j|^2\right>\\
   \nonumber
\left<r_\perp^2(s)\right> \equiv \sum_{i<j} \delta_{j-i , s} {1\over N-s} \left< |\Vr_{\perp i}-\Vr_{\perp j}|^2\right>
\end{eqnarray}

These two quantities are shown in fig. \ref{fig:rsq}, the upper and lower  curve being 
$\left<r^2(s)\right>$ and $\left<r_\perp^2(s)\right>$ respectively.
The best fit slope to $\left<r^2(s)\right>$ is $2\times .974$. The scaling
theory predicts a fractal dimension of $1$, which corresponds to a slope of $2$
on this plot.
$\left<r_\perp^2(s)\right>$ has a large plateau indicating an infinite fractal
dimension, consistent with our scaling analysis of chains in a tube-like 
conformations.

\begin{figure}
\centerline{\epsfxsize=\columnwidth \epsfbox{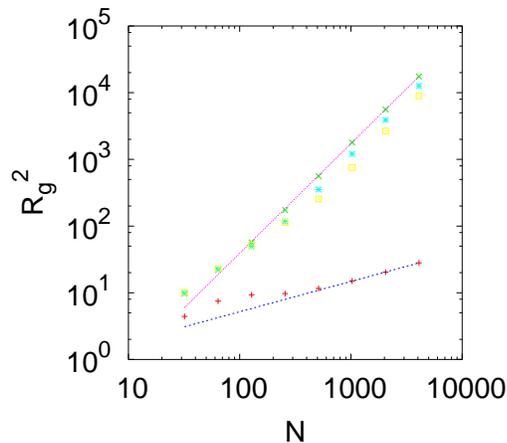}}
\caption
{
Radius of gyration squared versus chain length on a log-log plot for annealed
SAW's and vertical rods. The best fit
line is $2\times .82$ for the points at the highest rod filling fractions  (.32). 
The +'s show the radius of gyration in the x-y plane for the same conditions.
The other points show the radius of gyration at lower rod filling fractions.
}
\label{fig:annealed}
\end{figure}

Next we analyzed different chain lengths and rod densities.
The results are shown in figure \ref{fig:annealed}. The $\times$'s with the upper fitted
line going through them represent a fit of the exponent for chains of length $128$ to
$4096$ when the occupation fraction of rods is $\rho=.32$ . 
The fit gives an exponent of $.82$ which compares well with the
predicted value $.81$ given in eqn. (\ref{eq:rz}). The $*$'s and $\Box$'s are
the same data for filling fractions of $.16$ and $.08$ respectively. 

At all filling fractions there is a crossover from small chain lengths, which 
show the usual self avoiding behavior, to the elongated regime studied here.
From the figure it is clear that for small chain lengths the radius of gyration
becomes independent of rod filling fraction as expected.
It is also apparent from the figure that this crossover chain length decreases as
rod filling increases, which is in agreement with our theoretical interpretation
of the data, which predicts this cross-over chain length $N_c$ scales as 
$N_c \sim \rho^{-2\nu}$. 

The ``+" symbols with the straight line fit going through them
is the radius of gyration in the x-y plane. It is not
clear if this has  reached an asymptotic power law regime even at the larger chain lengths
because the slope of this line is quite small and therefore crossover effects
from short chains should be much larger than for $R_z$. The slope of the fit shown is $.23$
which is nonetheless not too far from the predicted value of $.27$.

\begin{figure}
\centerline{\epsfxsize=\columnwidth \epsfbox{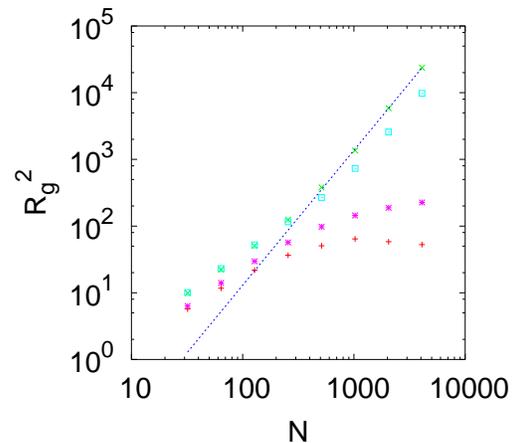}}
\caption
{
Radius of gyration squared versus chain length on a log-log plot for quenched
SAW's and vertical rods. The $\times$ and square symbols denote a rod filling
fraction of $.16$ and $.08$ respectively. The best fit
line is $2\times 1.01$ for the points at filling fractions of $.16$. 
The +'s and *'s show the radius of gyration in the x-y plane for the same conditions.
}
\label{fig:quenched}
\end{figure}

Now we turn to the results found when the rods are frozen to random positions.
In this case we did not go higher than a rod filling fraction $.16$ because
we wanted to stay far away from the percolation transition and the break down 
of ergodicity associated with it. The lattice size was $128\times128$ in the
x-y direction.
The chains were averaged over a large number of steps. The most time consuming
case were the longest chains, $N=4096$.
Before averaging, a chain was equilibrated for
$1.6\times 10^8$ steps and averaged over $1.28\times 10^8$ steps.
We then averaged the results over 128 realizations of the quenched rods. 
The radius of gyration squared is plotted
as a function of the number of monomers. The $\times$'s and the $\Box$'s
denote a rod filling fraction of $.16$ and $.08$ respectively. The slope
of the line going through the $.16$ filling fraction data gives $R_g \sim
N^{1.01}$, which is {\em higher} than the annealed case. On the other hand,
the radius of gyration in the x-y plane levels off as shown by the  two lower
curves. The +'s and the *'s are for a rod filling fraction of 0.16 and 0.08
respectively.  Here, in contrast to the annealed case---figure \ref{fig:annealed}---a leveling off in the curves is apparent.

The higher slope of $R_g(N)$ for quenched as opposed to annealed disorder
can be understood as follows. As discussed above, for large enough lattices
one expects the quenched average to equal the annealed one. The polymer
migrates around the lattice and finds rare locations where it will spend
most of its time. In these locations there will be a low rod density,
and the radius of the depleted region scales as in eqn. \ref{eq:rxy}. 
As the chain length grows, this implies that these regions must become
successively rarer because the size of these holes must grow in order to accommodate the chain. On the other
hand, if the lattice size is finite, then the strength of the most attractive
region is 
now bounded and cannot grow with N; hence the leveling off of the 
horizontal radius 
of gyration seen in figure \ref{fig:quenched}. A polymer chain in a fixed
tube has a overall radius of gyration $R_g \propto N$\cite{degennes} in good 
agreement with our findings.

\begin{figure}[ht]
\centerline{\epsfxsize=\columnwidth \epsfbox{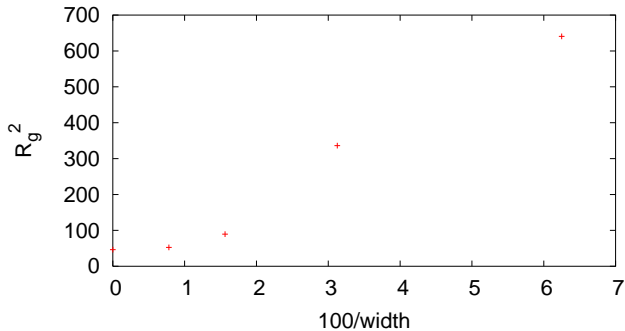}}
\caption
{
Horizontal radius of gyration squared versus horizontal system size. The
horizontal axis is 100 times the inverse width.
}
\label{fig:finitesize}
\end{figure}

To confirm this explanation, we ran the simulation for chains of length
$4096$ for different horizontal system sizes, $16\times16$ up to $128\times128$. 
The horizontal radius of gyration 
for different lattice sizes  is plotted in figure \ref{fig:finitesize}
as a function of inverse system size.  The point at zero represents the 
annealed cases which should be the same as the infinite lattice.
As the system width decreases, the horizontal radius of gyration {\em increases}
indicating that the polymer is less tightly bound to rod-depleted regions.

For many chain systems the rods will induce an attractive interaction 
between different chains. The annealed case is simplest to understand.
When two chains bind, they will stack on top of each other but
should not inter-penetrate as this strongly raises the free energy.
From the above scaling argument, eqn \ref{eq:rxy},
the free energy of
two chains binding should scale as the horizontal area occupied by a chain
$R_{xy}^2 \sim N^{2/(2+1/\nu)} \approx N^{0.54}$.  Therefore as polymer concentration
is increased from zero, one expects that
there will be a transition between a gas phase of isolated chains and a phase of chains
stacked on top of each other. 
Eventually it will reach another regime where one needs to consider the finite extent
of rods in the vertical direction. As the concentration is increased, at some
point different groups of stacked chains will want to coalesce. At this point
the system will demix into rod rich and polymer rich regions. The exact nature
of the phase diagram is probably quite complex and requires a detailed
understanding of the behavior of
semi-dilute polymer solutions in pores~\cite{brocharddegennes,degennes}, which
itself is quite delicate.

This is also hard to simulate quantitatively (even in the annealed
case) because the center of mass diffusion of polymers is very
slow on a monte-carlo time scale. This means that for a large
system volume, it takes a long time for two chains to coalesce.
However, it was observed that in the pillar regime, two chains that 
finally managed to diffuse into the same tube stayed bound to each
other.

These results found here differ strongly from those found for a model for an SAW
in the presence of a solution of vertically aligned short rods~\cite{vliet}
where their SAW's were of order 100 units but the rods were only
one or two lattice spacings. 
In that case, the authors found that the polymers went into disk-like
configurations instead of the rod-like ones found here. 
Here it is worthwhile examining the case of rods of length $L$ and
chains whose vertical extent is much larger. In this case 
eqn. \ref{eq:anneal} implies an attractive potential
between monomers with the same x-y coordinates whose 
vertical separation is less than $L$. At separations larger than $L$
we expect that chain segments will no longer be stacked in the
same tube because of the absence of attraction at this
length scale. 
This, according to eqn \ref{eq:rz}, corresponds to a crossover chain length of $N_c \sim L^{(2+1/\nu)/3}$
Beyond this separation, different monomers become unbound and
we expect that in the horizontal direction segments execute a random walk. 
Therefore for large $N >> N_c$, 
$R_{xy} \sim  L^{1/3} (N/N_c)^{1/2} \sim  L^{1/3} L^{-(1+1/(2\nu))/3} N^{1/2} \sim L^{-1/(6\nu)} N^{1/2}$. 
Therefore we expect a slow drift of the SAW in the vertical direction that
decreases with increasing $L$.

In conclusion, we have shown that a polymer in the presence of either mobile or
frozen vertical rods forms tube-like vertical conformations where the tube
diameter and tube length scale as in eqns \ref{eq:rxy} and \ref{eq:rz}. It
should be possible to observe this unusual behavior experimentally. One
possible system for the case of frozen rods would be in forests of vertical
nanotubes~\cite{semet} suitably coated to make them repulsive to polymer chains\cite{nanotubes}.
These results are also relevant to the more complicated case of polymer-rod
mixtures, which are usually analyzed assuming no change in polymer statistics
due to the presence of rods.


\begin{references}
\bibitem{dimarzio} E.A. Di Marzio, Phys. Rev. Lett. {\bf 64} 2791 (1990).
\bibitem{baumgartner} A. Baumgartner and M. Moon, Europhys. Lett. {\bf 9} 203
(1989).
\bibitem{nelson} D.R. Nelson, Phys. Rev. Lett. {\bf 60}, 1973 (1988).
\bibitem{nattermann} T. Nattermann and R. Lipowsky, Phys. Rev. Lett. {\bf 61}
2508 (1988).
\bibitem{kardar} M. Kardar and Y.-C. Zhang, Phys. Rev. Lett. {\bf 58}, 2087
(1987).
\bibitem{arsenin} I. Arsenin, T. Halpin-Healy, and J. Krug, Phys. Rev. E {\bf 49} 
R3562 (1994).


\bibitem{nanotubes} M.F. Islam MF, E. Rojas, D.M. Bergey,  A.T. Johnson and A.G. Yodh, Nano Lett. {\bf 3} 269 (2003).


\bibitem{dogic} Z. Dogic, K. R. Purdy, E.  Grelet,  M. Adams and S. Fraden, Phys Rev E {\bf 69} 051702 (2004).
\bibitem{sear} R.P. Sear, J. Phys. {\it II France} {\bf 7} 877 (1997). 
\bibitem{buitenhuis} J. Buitenhuis, L. N. Donselaar, P. A. Buining, A. Stroobants, and H. N. Lekkerkerker, 
J. Colloid Interface Sci. {\bf 175}, 46 (1995). 
\bibitem{vanbruggen} M. P. B. van Bruggen and H. N. W. Lekkerkerker, Macromolecules {\bf 33}, 5532 (2000). 
\bibitem{edgar} C. D. Edgar and D. G. Gray, Macromolecules {\bf 35}, 7400 (2002). 
\bibitem{tang} J. X. Tang, S. Wong, P. T. Tran, and P. A. Janmey, Ber. Bunsenges. Phys. Chem. {\bf 100}, 796 (1996). 
\bibitem{devries} R. de Vries, Biophys. J. {\bf 80}, 1186 (2001). 
\bibitem{semet} V. Semet, Vu Thien Binh, P. Vincent, D. Guillot, M. Chhowalla, 
G. A. J. Amaratunga, W. I. Milne, P. Legagneux and D. Pribat, 
Appl. Phys. Lett. {\bf 81} 343 (2002)
\bibitem{catesball} M. E. Cates and C. Ball, J. Phys. (France) {\bf 49}, 2009 (1988).
\bibitem{machta} J. Machta and R.A. Guyer, J. Phys. A {\bf 22} 2539 (1989).
\bibitem{lee} S.B. Lee and H. Nakanishi, Phys. Rev. Lett. {\bf 61}, 2022 (1988).
\bibitem{degennes} P.G. de Gennes "Scaling Concepts in Polymer Physics" Cornell University Press (1985).
\bibitem{wallmandell} F.T. Wall and F Mandell, J. Chem. Phys. {\bf 63} 4592 (1975).
\bibitem{brocharddegennes} F.Brochard and P.G. de Gennes, J. Phys. Lett.  (France) {\bf 40} 399 (1979).
\bibitem{vliet} J.H. van Vliet, M.C. Luyten and G. ten Brinke, J. Phys. II (France) 603 (1993).
\end{references}
\end{document}